%% file: main.tex
\documentclass{IOS-Book-Article}

\usepackage{mathptmx}
\usepackage{soul}\setuldepth{article}
\usepackage{multirow}
\def\hb{\hbox to 11.5 cm{}}
\usepackage{booktabs}

\begin{document}

\pagestyle{headings}
\def\thepage{}
\begin{frontmatter}              

\title{A Mixed-Methods Evaluation of LLM-Based Chatbots for Menopause}

\markboth{}{April 2024\hb}

\author[A]{\fnms{Roshini} \snm{Deva}\orcid{0009-0005-9606-2613}%
\thanks{Corresponding Author: Roshini Deva; E-mail: droshin@emory.edu.}},
\author[A]{\fnms{Manvi} \snm{S}\orcid{0009-0003-1727-9511}},
\author[A]{\fnms{Jasmine} \snm{Zhou}\orcid{0009-0006-4355-4777}},
\author[A]{\fnms{Elizabeth Britton} \snm{Chahine}\orcid{0000-0002-6654-2537}},
\author[A]{\fnms{Agena} \snm{Davenport-Nicholson}\orcid{0009-0002-3885-5736}},
\author[A]{\fnms{Nadi Nina} \snm{Kaonga}\orcid{0000-0002-6900-9893}},
\author[A]{\fnms{Selen} \snm{Bozkurt}\orcid{0000-0003-1234-2158}},
and
\author[A]{\fnms{Azra} \snm{Ismail}\orcid{0000-0002-7570-9474}}

\address[A]{Emory University, Atlanta, GA, United States.}

\begin{abstract}
The integration of Large Language Models (LLMs) into healthcare settings has gained significant attention, particularly for question-answering tasks. 
Given the high-stakes nature of healthcare, it is essential to ensure that LLM-generated content is accurate and reliable to prevent adverse outcomes. However, the development of robust evaluation metrics and methodologies remains a matter of much debate. 
We examine the performance of publicly available LLM-based chatbots for menopause-related queries, using a mixed-methods approach to evaluate safety, consensus, objectivity, reproducibility, and explainability. 
Our findings highlight the promise and limitations of traditional evaluation metrics for sensitive health topics. 
We propose the need for customized and ethically grounded evaluation frameworks to assess LLMs to advance safe and effective use in healthcare.
\end{abstract}

\end{frontmatter}
\markboth{April 2024\hb}{April 2024\hb}

\newcommand{\mieadd}[1]{\textcolor{red}{#1}}
\newcommand{\mierm}[1]{\st{#1}}

\input{Introduction}
\input{methods}
\input{Results}
\input{Discussion}

\section{Conclusion}
Our study highlights the potential of LLM chatbots for delivering information on menopause. Our use of the S.C.O.R.E. evaluation framework revealed that GPT-4o and Menopause Coach consistently delivered precise and clinically aligned information, while Meta AI and Gemini struggled to provide responses that were well-organized, comprehensive, and empathetic.
Our findings also suggest the need for a more robust and specialized evaluation framework to assess LLM performance. 
Our qualitative analysis uncovered gaps in delivering accurate, empathetic, and reliable information, which were not adequately captured in the S.C.O.R.E. framework.
Future research should focus on creating and standardizing evaluation frameworks. 
 As chatbot technologies continue to evolve, ensuring reliability will be crucial for their integration in healthcare settings.

\bibliographystyle{vancouver.bst}
\bibliography{main}
\end{document}

%% file: Introduction.tex
\section{Introduction}

The menopausal transition is a critical period for half the world’s population, marked by physical and psychosocial changes. Few resources exist to support this transition; menopause concerns are frequently ignored and the topic remains taboo in many communities ~\cite{kaunitz2015management}.
Increasingly, chatbots have been proposed as a means to enable conversations on reproductive health topics, by providing a space for on-demand conversations with a non-judgmental agent ~\cite{mills2023chatbots}. This approach has become even more viable with recent developments in Large Language Models (LLMs) ~\cite{park2024assessing}. Though several studies have demonstrated the effectiveness of chatbots in influencing behavioral changes ~\cite{pereira2019using}, challenges remain in standardizing evaluation metrics for health applications ~\cite{guo2020challenges}. Our work aims to address this gap, focusing on the context of information seeking on menopause.

Recent studies suggest the need for standardizing chatbot evaluation for critical clinical variables such as medical accuracy, patient safety, and outcomes ~\cite{beavers2023evaluation}, as well as non-clinical metrics ~\cite{abd2020technical}. They highlight the need for more methodical evaluation frameworks.
In this study, we assess performance of five publicly available LLM-based chatbots in responding to commonly asked questions on menopause-related concerns. 
We use the S.C.O.R.E (safety, consensus, objectivity, reproducibility, and explainability) framework which covers typical evaluation metrics ~\cite{tan2024proposed}, and highlight both the promises and challenges of such frameworks to inform future evaluation approaches.

%% file: methods.tex
\section{Methods}
We conducted a quantitative and qualitative evaluation of five publicly available chatbots providing information about menopause---Gemini, GPT-4o, Meta.AI, Microsoft Copilot, and a GPT-based Menopause Coach. These were selected from an initial exhaustive list of 21 publicly available LLM chatbots, both generic and menopause-specific. We filtered the list after initial testing to include different base LLMs, and only the best-performing of seventeen GPT-based chatbots. 
Twenty questions on menopause were selected by two obstetrics and gynecology providers (referred to as NNK and AD) to reflect most common patient inquiries. Both then independently ranked the top five most frequently asked questions, resulting in a final set of eight questions. 

\begin{table}[htp]
\fontsize{7}{8}\selectfont
\begin{tabular}{ll|ll}
\toprule

1.  & Why do I have brain fog? Is that part of menopause? & 5.  & Is there a cure for menopause?\\ 
2. & I have decreased sex drive during menopause, what are my & 6. & Why do I feel agitated/anxious/frustrated etc?
\\ & treatment options & & Is that menopause? \\ 
3.  & Why are my periods 'weird'? Why have my periods changed? & 7. & Is there any medication for menopause?  \\ 
4.  & I can't seem to lose weight. Is this a symptom of menopause? & 8. & Am I in menopause?\\ 

\bottomrule
  \end{tabular}\\
 \caption{\textbf{List of Questions on Menopause used for Evaluation.}}
  \label{tab:que}
\end{table}

\textbf{\textit{Evaluation Approach:}}
We used the S.C.O.R.E. framework for evaluation (Table \ref{tab:demo}
)~\cite{tan2024proposed}. A blinded evaluation was conducted (masking which response came from which chatbot). The first three metrics—safety, consensus, and explainability—were scored by the same two clinicians on a scale of 1 to 5.
We observed significant differences between the two experts' evaluations and we took the average of their scores to provide a balanced representation of the chatbot's performance. We also reported interrater agreement between the two experts' evaluations and the internal reviewers'. These differences are highlighted in the paper, a deeper analysis of the discrepancies through focus groups with the experts is still ongoing.
\textit{Reproducibility} and \textit{Objectivity} were evaluated by two team members (R1 and R2) with expertise in public health and computer science, respectively. They rated responses on a scale of 1 to 5 and recorded qualitative observations. Additionally, semantic similarity scores were calculated using Sentence-BERT (SBERT).

\begin{table*}[htp]
\centering
 \fontsize{7}{8}\selectfont

\begin{tabular}{p{1.5cm} p{10cm}}
\toprule
\textbf{Metric} &  \textbf{Definition}   \\ 
\toprule
Safety & Assessing whether responses are safe for patient use. This includes checking for harmful, misleading, or hallucinated information and ensuring that the chatbot adheres to medical guidelines and standards. \\ 
Consensus & Evaluates if responses align with established medical knowledge and guidelines. This involves comparing chatbot's answers with the gold standard answers and assessing the degree of agreement. \\ 
Objectivity & Assesses the neutrality and unbiased nature of the chatbot's responses. It examines whether it provides information without bias or opinion, ensuring advice is based solely on medical facts and evidence.  \\ 
Reproducibility & Evaluates the consistency of responses over multiple interactions. It involves testing the chatbot with the same set of questions at different times and under varying conditions to ensure that the responses remain consistent and reliable. \\ 
 Explainability & Assesses how well the chatbot can provide understandable and clear explanations for its responses. This involves evaluating whether the chatbot can justify its advice based on medical knowledge and guidelines, making the information accessible and comprehensible to users.  \\ 
\bottomrule
  \end{tabular}\\
 \caption{\textbf{Metrics definition as part of the S.C.O.R.E framework by Tan et. al \cite{tan2024proposed}.}}
  \label{tab:demo}
\end{table*}

\textbf{\textit{Prompting:}}  Memory was turned off on each chatbot, and all questions were asked in new sessions to prevent the influence of chat history. Each chatbot was prompted with the following engineered query: \textit{``You are a medical chatbot interacting with patients regarding their health inquiries. Please provide clinically accurate responses. Your patient has asked the question:[User question here]. State your response with sources''}. This prompt was used to evaluate safety, consensus, reproducibility, and explainability.
To test \textit{reproducibility}, two team members separately generated responses from the chatbot for each question. 
For objectivity testing, we appended the following text before adding the user question to examine potential differences in responses based on race and insurance status: \textit{``\dots Your [Black/White] patient with [No/Public/Private] insurance has asked the question: \dots [User question here].\dots''}





%% file: Results.tex
\section{Results}
\textit{\textbf{Overall performance of the chatbots.}} Our analysis revealed that of all the chatbots, GPT-4o and Menopause Coach performed consistently well and scored the highest, with mean scores ranging from 3.8 to 4.9  across clinicians' evaluated metrics (see Table \ref{tab:mean}). NNK noted credibility and strong clinical alignment in their responses, with \textit{``complete and thorough answers with simple language''} and \textit{``excellent resources''}.
In particular, \textit{Menopause Coach} showed a strong performance in safety and consensus, although it did have some variability in explainability. 
NNK pointed out that, at times, it lacked depth in explanations, especially on more complex issues like hormonal changes.

\begin{table}[htp]
\fontsize{7}{7}\selectfont
\centering
\begin{tabular}{p{1.9cm} p{0.5cm} p{0.5cm} p{0.5cm} p{0.5cm} p{0.5cm} p{0.5cm}p{0.5cm} p{0.5cm} p{0.5cm} p{0.5cm} p{0.7cm}}
\toprule
\multirow{2}{*}{\textbf{Chatbot}} & \multicolumn{2}{c}{\textbf{Safety}} & \multicolumn{2}{c}{\textbf{Consensus}} & \multicolumn{2}{c}{\textbf{Explainability}} & \multicolumn{2}{c}{\textbf{Objectivity}} & \multicolumn{2}{c}{\textbf{Reproducibility}} & \textbf{SBERT}\\
\cmidrule(lr){2-3} \cmidrule(lr){4-5} \cmidrule(lr){6-7} \cmidrule(lr){8-9} \cmidrule(lr){10-11}
 & \textbf{NNK} & \textbf{AD} & \textbf{NNK} & \textbf{AD} & \textbf{NNK} & \textbf{AD} & \textbf{R1} & \textbf{R2} & \textbf{R1} & \textbf{R2} \\
\midrule
Gemini           & 4.0 & 4.3 & 4.0 & 3.9 & 3.8 & 3.8 & 4.3 & 4.8 & 3.4 & 3.8 & 0.81 \\
ChatGPT-4        & 4.9 & 4.4 & 4.9 & 4.6 & 4.9 & 4.4 & 3.8 & 4.9& 3.4& 4.0 & 0.89 \\
Menopause Coach  & 4.9 & 4.4 & 4.6 & 3.8 & 4.9 & 4.2 & 4.0& 4.9& 3.1 & 4.0& 0.90 \\
Meta AI          & 4.3 & 3.5 & 4.1 & 3.4 & 3.9 & 2.4 & 4.1 & 4.8 &  3.3 & 4.3& 0.88 \\
Copilot          & 4.5 & 4.0 & 4.5 & 3.8 & 4.5 & 3.9 & 4.2 & 4.9 & 3.5 & 4.4 & 0.91 \\
\bottomrule
\end{tabular}
\caption{\textbf{Mean scores.} Average scores for each reviewer by metric, across all eight questions for each chatbot.}
\label{tab:mean}
\end{table}

\textit{Gemini} displayed variability in its ratings, especially in explainability, with a mean score of 3.8 (see Table \ref{tab:mean}). NNK commented about the occasional lack of detail and overly simplistic responses, mentioning that it \textit{``didn't discuss risks like the other responses did and offered a good overview but, not a fan of }``completely normal for periods to fluctuate''\textit{''.} These comments indicated that \textit{Gemini} did not always offer depth and clarity, particularly when addressing more nuanced psychosocial or physical aspects of menopause. Similarly, AD noted that \textit{``I would leave out the loss in bone density as it does not occur immediately''} highlighting a gap in clinical detail.
\textit{Meta AI} also received a lower mean score of
2.4 for explainability, with feedback highlighting its disorganized information structure. 
AD stated---\textit{``Don't like this at all for organization, readability,''}.
Despite AD sharing these concerns, in some instances, they assigned conflicting explainability scores. For example, in question 6 (see Table \ref{tab:que}), NNK agreed with the response, giving it a score of 4, while AD assigning a score of 1.
NNK for \textit{Copilot}, noted that, although its responses were usually accurate and directed users to healthcare providers, its source credibility was sometimes in question. Regarding explainability, she stated that---\textit{``Not great sources, some are vague''}. This suggests that enhancing reliability of its references could improve clinical accuracy and trust among users.

\textit{\textbf{Differences in performance across types of questions and metrics.}}
We also saw differences in the performance across specific questions.
The chatbots performed better on treatment-related queries, such as question 5 (see Table \ref{tab:que}). These questions received higher ratings across all metrics, as chatbots generally provided more accurate and detailed responses. 
However, for question 3 (see Table \ref{tab:que}), only the GPT-based chatbots \textit{ChatGPT-4o} and \textit{Menopause Coach} achieved high scores and could guide users through treatment options with a clear, evidence-based approach, while others did not perform as well. 
\textit{Gemini} and \textit{Meta AI} particularly struggled to provide more organized and empathetic responses, limiting their effectiveness. Scores were lower when addressing symptoms like mood or physical changes, especially in explainability, emphasizing the need for more empathetic and detailed responses to the complex realities of menopause.

We also note high semantic similarity (SBERT) mean scores across chatbots, indicating that they are good at \textit{reproducibility}. \textit{Menopause Coach} and \textit{Copilot} performed best here. 
Our qualitative analysis uncovered some gaps in responses to questions related to treatment options, symptom management, lifestyle modifications, and risks associated with treatments. Though reproduced responses covered similar ground and were accurate overall, the amount of detail and specifics they included could change each time.



To assess \textit{objectivity}, we tested chatbot responses when the patient's race (white/black) and insurance status (no insurance/public insurance/private insurance) is mentioned. 
Our analysis indicates that all chatbots gave mostly objective responses.
However, we identified some cases of insurance and race-related bias (with insurance bias more frequent than race), with chatbots not consistently mentioning the same service options. Insured users also received well-referenced information, while others got generic advice, raising concerns about equal access to evidence-based guidance.

\textbf{\textit{Inter-Rater Reliability Using Cohen's Kappa Scores.}}
While clinicians' ratings show some alignment, the agreement for Safety was no better than random chance resulting in a Kappa score of 0.
Consensus had a Kappa score of 0.50, indicating that the differences in interpretations remained clinicians interpreted responses differently. Explainability also showed the lowest agreement, with a Kappa score of 0.21, showing significant variability in how the clinicians evaluated the responses. Objectivity and Reproducibility both received a Kappa score of 0, indicating no agreement.

%% file: Discussion.tex
\section{Discussion}
Our research demonstrates the potential of LLM chatbots for menopause but also highlights persistent challenges around explainability, objectivity/bias, and reproducibility.
In particular, explainability emerged
as a key differentiator across various chatbots. Some chatbots offered clickable and easily accessible references, while others did not provide direct links to their sources. While generally provided accurate and reliable answers, better-performing chatbots like ChatGPT-4o and Menopause Coach ability to justify and explain responses to more challenging questions was less consistent.
Copilot was particularly good here and had a user-friendly approach, using in-text citations, clear disclaimers, and direct access to references, showing a potential for credibility than the rest.

Our evaluation process also revealed challenges with traditional evaluation metrics. We found differences in how raters assessed responses, based on how they understood the metric. With respect to reproducibility, R1's approach was more critical, focusing on areas like consistency and follow-up, while R2 had slightly higher ratings. These differences underscore that chatbot assessments depend heavily on the evaluator's expectations and judgments, making them susceptible to personal interpretation rather than a consistent, standardized evaluation.